# Optimization of Bluetooth Audio Stream based on the Estimation of Proximity

Ka. Selvaradjou, A. Sharma Shankar, U. Anandakumar and N. Sivasundar

*Abstract*— **The advent of Bluetooth wireless technology makes it possible to transmit real-time audio in mobile devices. Bluetooth is cost-efficient and power-efficient, but it is not suitable for traditional audio encoding and real-time streaming due to limited bandwidth, high degree of error rates, and the time-varying nature of the radio link. Therefore, audio streaming over Bluetooth poses problems such as guzzling of both power and bandwidth. In order to overcome the above mentioned problems, an algorithm is proposed in this work to optimize the audio stream from the source to the sink by estimating the proximity between them. The optimization is achieved by adjusting the bit rate of the audio stream thus conserving power. We considered carefully various Bluetooth signal parameters and the most suitable parameter for estimating the proximity has been determined experimentally. The experiments were carried out using Class II BS003 Bluesoleil dongle. This work will enable the Bluetooth users to perform a seamless and optimized streaming of MP3 stereo audio data.**

## I. INTRODUCTION

The recent growth of home networking applications has raised the requirement for cable-free distribution of an increasing amount of media to portable playback devices. Technology has arrived that can deliver wireless access to multimedia information such as hypertext, audio, images, video, and 3D graphics. Since the primary concern of any wearable playback device is the limited amount of battery energy available, it is important to understand the trade-offs involved in designing and configuring a wearable audio playback device. The components that account for most of the energy consumed in a Bluetooth playback device are (a) receiving the compressed audio over Bluetooth, (b) decoding the compressed audio and (c) outputting the analog audio signal. Decoding a compressed audio with high bitrate requires a large processing power when compared to audio with low bitrate. Moreover the quality of the Bluetooth link depends on the distance from the audio server. If the quality of the wireless link is very poor, with whatever bitrate the audio is streamed, the quality of the audio is going to be poor. So, the battery power is wasted by streaming with high bitrate at long distances. Here, the optimization is carried out by conserving the power during audio streaming by calculating the proximity of the listener and appropriately changing the bit rate of the audio stream. Moreover optimization can be extended by incorporating a warning system intimating a music-listener if he goes beyond the range of the transmitted signal, thus preventing an abrupt loss of music. This could be achieved by continuously estimating the proximity of the listener with the server. If the audio server is powerful enough, in terms of computational and memory resources, it can save the session of the audio client to resume later or can execute an automated process.

Besides audio applications, the estimated proximity can be used to warn Bluetooth gamers if they are on the verge of disconnection. The estimated proximity can also be used for location and/or positioning systems enabling Bluetooth users to be tracked and guided automatically. Rest of the paper is organized as follows. Chapter 2 discusses the related work. In Chapter 3, we discuss the various proximity estimation techniques. Chapter 4 discusses the audio streaming between the Bluetooth devices and its related issues. In Chapter 5, the experimental results are presented and discussed and in Chapter 6 the work is concluded and future directions for this work are presented.

## II. LITERATURE REVIEW

Several methods were proposed for Bluetooth positioning such as Time of Arrival (TOA) [1], Time Difference of Arrival (TDOA) [1] and Angle of Arrival (AOA) [1]. TOA is based on triggering the mobile devices to respond, and measuring the time it takes for the response to reach back to the antenna. The elapsed time is used to measure the distance between the two. Time difference of Arrival technique uses a combination of Radio Frequency (RF) and Ultrasonic to enable a listener to determine the distance to beacons, from which the closest beacon can be unambiguously inferred. AOA is based on finding the direction of maximum signal intensity for each antenna-device pair. By finding the intersection of few such direction vectors, a mobile's position can be estimated. AOA is considerably less accurate than TOA, due to angular resolution and the fact that in indoor environment much of the signal is reflected. Positioning was later done based on signal strength using Triangulation. Triangulation [2] is based on measuring the Received Signal Strength Indication (RSSI) from each antenna (possibly by using triangulation mechanism), with respect to each mobile device. Positioning was later made accurate by using a more accurate method namely least squares estimation [3]. Preliminary Audio Streaming was done at radio quality (22.050 KHz, 8 bit, mono) [4]. This was done using GOEP (Generic Object Access Profile) and Generic Audio Profile (GAP). It can basically be considered as a Synchronous Connection Oriented (SCO) link. Audio Streaming supports both asynchronous and isochronous services through two different types of links: (a) point-to-point or point-to-multipoint Asynchronous Connection-Less (ACL) links and (b) Synchronous Connection Oriented point-to-point links. When using an ACL link, a "slotted" channel is applied with a slot time equal to 625μs, while a Time-Division Duplex (TDD) scheme is employed for full duplex transmissions. A packet transmission can cover up to 5 channel slots. The packet-oriented transmissions through the ACL link scan achieve a maximum of 721kbps effective bit rate and can be acknowledged through a retransmission mechanism [5]. To achieve streaming at higher bit rate (higher audio quality) streaming was done later using ACL links.

### III. PROXIMITY ESTIMATION

The following are the parameters taken into account for proximity estimation in this study:

- Received Signal Strength Indication (RSSI)
- Round Trip Time (RTT)
- Inquiry based RSSI
- Link Quality (LQ)

#### A. Received Signal Strength Indication (RSSI)

In the IEEE 802.11 systems [6], the RSSI is the relative received signal strength in a wireless environment. RSSI is an 8-bit signed integer that denotes whether the received (RX) power level is within or above/below the Golden Receiver Power Range (GRPR) [7]. The GRPR is regarded as the ideal RX power range. Fig. 1 illustrates the relationship between RSSI and GRPR as defined in Bluetooth specification. A positive RSSI indicates that the RX power level is above GRPR, whereas the negative RSSI indicates that the RX power level is below GRPR and a zero RSSI implies that RX power level is ideal (i.e., within GRPR). The lower and upper thresholds of GRPR are loosely bound, leaving them to be device specific, which in turn affects the RSSI, since it is merely a relative parameter.

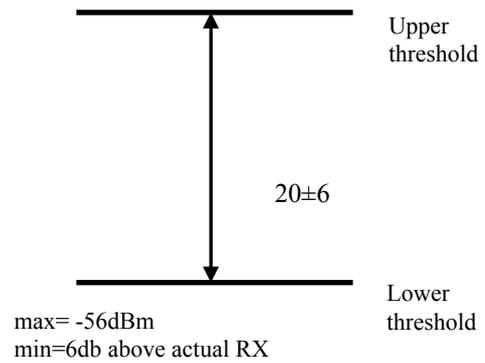

Upper threshold

20±6

Lower threshold

max= -56dBm
min=6db above actual RX

Fig. 1 Golden Receiver Power Rank of RSSI

#### B. Round Trip Time (RTT)

The term round-trip delay time or round-trip time (RTT) is the elapsed time for transit of a signal over a closed circuit, or time elapsed for a message. A packet is setup and it is sent to the receiver and the receiver again sends back the packet spontaneously to the sender. The connection is established across the L2CAP (Logical Link Control and Adaptation Protocol) [7]. The packets transferred across the L2CAP have a maximum size. This is called Maximum Transmission Unit (MTU). The packet size used was of the MTU size. It is also necessary that the client sends back the packet immediately because whenever packets are transferred from the server to the client, the client asks for the authentication of the user in the client side and then only accepts the packet which induces a time delay resulting in an erroneous RTT. The time delay due to authentication can be eliminated by pairing the devices. So, paring of devices should be done before the estimation of RTT.

#### C. Inquiry based RSSI

Inquiry based RSSI works in a similar manner as a typical inquiry. In addition to the other parameters (e.g., Bluetooth device address, clock offset) generally retrieved by a normal inquiry procedure, it also provides the RSSI value. Since it requires no active connection, the radio layer simply monitors the RX power level of the current inquiry response from a nearby device, and infers the corresponding RSSI. The Bluetooth inquiry procedure uses 32 dedicated inquiry hop frequencies (in countries with 79 Bluetooth frequency channels) according to the inquiry hopping sequence as defined in the Bluetooth specification [7]. The inquiry-hopping rate is twice the nominal frequency-hopping rate used by ordinary connections. In other words, an inquiring device switches to a new frequency every 312.5 μs, whereas a typical Bluetooth time slot is 625 μs long. The inquiry hopping sequence is split into two trains, A and B, of 16 frequencies each (see Fig. 2). In one slot (i.e., 625 μs), the inquiring device sequentially transmits on two different frequencies. In the following slot, it shall listen for any response to the previous two frequency hops, in the same sequence. Consequently, each train comprises 16 alternate transmitting and listening slots, and spans 625 μs × 16 = 10 ms. According to Bluetooth specification, a single train is repeated for at least 256 times before switching to a new train.

In an error-free environment, a Bluetooth device is recommended to perform at least three such switches in order to collect all responses. As a result, the whole inquiry procedure may last for $4 \times (256 \times 10 \text{ ms}) = 10.24$ sec, which can be a major drawback if latency is a prime concern.

### D. Link quality (LQ)

LQ is an 8-bit unsigned integer that evaluates the perceived link quality at the receiver. It ranges from 0 to 255; the larger the value, the better the link's state. For most Bluetooth modules, it is derived from the average bit error rate (BER) (Average BER is the number of error bits divided by the number of transmitted bits) seen at the receiver, and is constantly updated as packets are received. LQ is used mainly for adapting to changes in the link's state of the Bluetooth connection.

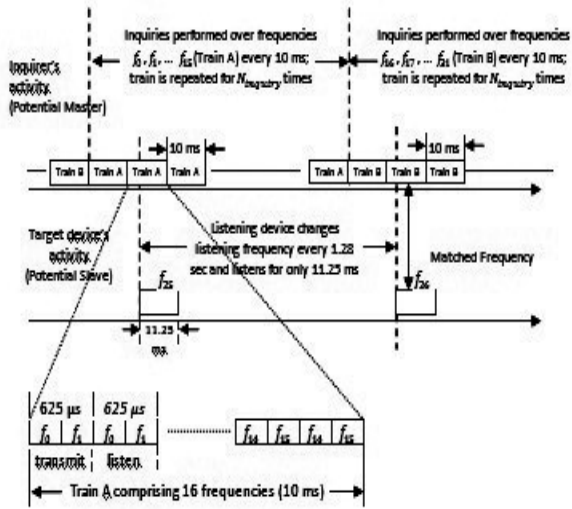

Fig.2 Potential Master's and Slave's frequency scanning during Bluetooth inquiry procedure (in countries with 79 Bluetooth frequency channels)

### IV. AUDIO STREAMING

Audio streaming is done by exploiting the A2DP (Advanced Audio Distribution Profile) of the Bluetooth stack [8]. This profile defines the requirements for Bluetooth devices necessary for supporting the high quality audio distribution. The Advanced Audio Distribution Profile (A2DP) defines the protocols and procedures that realize distribution of audio content of high-quality in mono or stereo on Asynchronous Connection-less (ACL) channels. The term 'Advanced Audio', therefore, should be distinguished from 'Bluetooth audio', which indicates distribution of narrow band voice on Synchronous Connection-Oriented (SCO) channels. A typical usage case is the streaming of music content from a stereo music player to headphones or speakers. The audio data is compressed in a proper format for efficient use of the limited bandwidth. When a device wishes to start streaming of audio content, the device firstly needs to set up a streaming connection. Before starting the streaming, the devices undergo Signalling procedure [8]. During such set up procedure, the devices select the most suitable audio streaming parameters. Once streaming connection is established, *Start Streaming* procedure in Generic Audio Video Distribution Profile

(GAVDP)[9] is executed, both source (SRC) and sink (SNK) are in the STREAMING state, in which the SRC (SNK) is ready to send (receive) audio stream. The SRC uses the *Send Audio Stream* procedure to send audio data to the SNK. The SNK in turn employs the *Receive Audio Str*eam procedure to receive the audio data. The devices shall be in the STREAMING state to send/receive audio stream. The media player is designed using GStreamer [10].

### V. RESULTS AND DISCUSSION

In this section, the results of the experiments that are carried out to compare the various parameters available for proximity estimation are presented. The experiments were conducted using the BlueZ stack [11] in Linux.

### A. Distance vs. RSSI

It can be observed from Fig. 3 that connectionless RSSI decreases with increase in distance between the sender and receiver, but it can't be used at the time of streaming, because no stable connection is available.

It can be observed from Fig. 4 that connection based RSSI initially decreases with increase in distance between the sender and receiver, and then it remains constant throughout the connection. RSSI denotes the RX power level. During a stable connection, connection based RSSI stays at a constant value when the receiver moves away from the transmitter. This is due to a feedback mechanism [7] that is incorporated in the Link Manager Protocol of the Bluetooth Hardware to ensure that the transmitted power is in the Golden Receiver Power Range (GRPR). Since this mechanism is built into the hardware and requires hardware level switches to disable it, this parameter also can't used. Hence RSSI also remains constant during a stable connection as shown in Fig. 4.

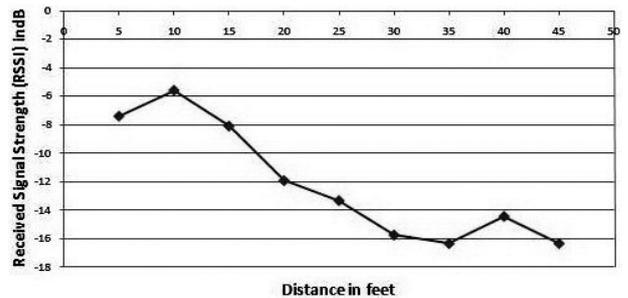

Fig.3 Distance vs. Connection-less RSSI

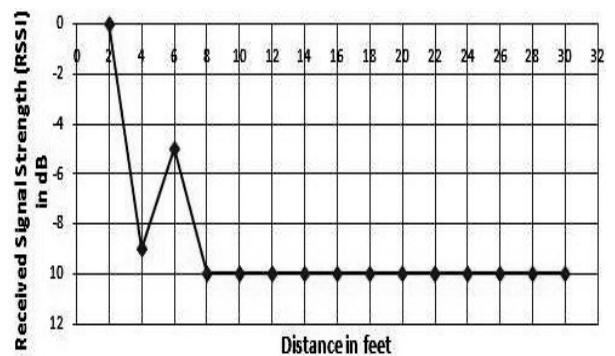

Fig.4 Distance vs. Connection-based RSSI

### B. Distance vs. Round Trip Time

The results plotted in Fig. 5 show the variation of RTT with distance. A packet of fixed size is sent to the receiver and it is again got back in the server. The client is coded in such a way that it sends the packet immediately without any acknowledgement. As it can be seen RTT varies linearly with distance. RTT cannot be used to estimate proximity because RTT depends on several other parameters other than distance such as number of devices in the piconet and Forward Error Correction packets. Moreover, RTT packets consume considerable bandwidth in the estimation of the proximity and the RTT [Fig. 5] itself is too large at the far end (point of disconnection). Hence RTT is a poor candidate for proximity estimation during audio-streaming.

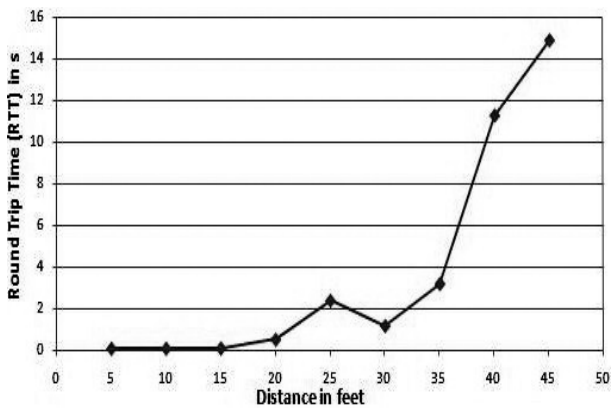

Fig.5 Distance vs. RTT

### C. Distance vs. Inquiry based RSSI

Fig. 6 shows that Inquiry based RSSI does not linearly decrease with increase in distance between the sender and receiver. Though Inquiry based RSSI works free from the feedback mechanism that is incorporated in the Link Manager Protocol of the Bluetooth Hardware and might seem like the best-suited parameter, this requires the client to in inquiry mode even when there exists a connection (when streaming audio). This mode of operation is not supported by any standard Bluetooth headphone available in the market and thus this also cannot be used in the estimation of proximity.

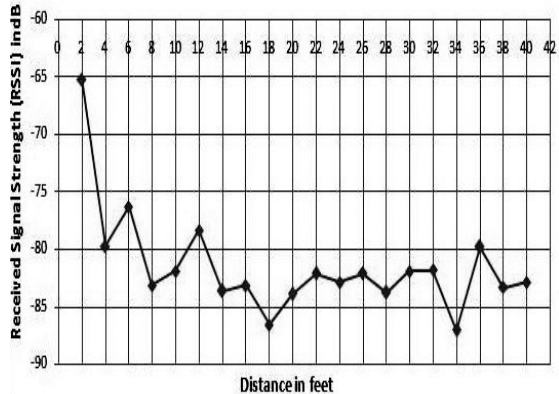

Fig.6 Distance vs. Inquiry based RSSI

### D. Distance vs. Link quality

Link Quality [Fig. 7] has a relatively better relation with distance than the other available parameters although LQ readings obtained at smaller distances show a little variation. Also Link quality is a connection-oriented parameter which makes it suitable to be used during audio streaming. When compared to other parameters that are considered for proximity estimation, LQ correlates much better with the distance.Thus, Link quality can be used to estimate the proximity of the receiver from the sender while audio streaming.Eventhough the experiments are carried out using a Class II Bluetooth dongle, the characteristics of Link quality were taken till the end of total disconnection. (i.e. not even a single packet is received at the receiver's end). The irregular variations at the end of disconnection are due to the feedback mechanism [7]. It can be observed from fig. 7 that the variation of Link quality with distance till the range of 30 feet is almost uniform. The link quality is recorded and if the the link quality seems to increase then it means that the receiver comes closer to the sender and if the link quality seems to decrease then it means that the receiver goes away from the sender. Thus, the proximity is determined and if the receiver goes away from the sender, the bitrate of the audio is decreased and if the receiver comes closer to the sender, the bitrate of the audio is increased.

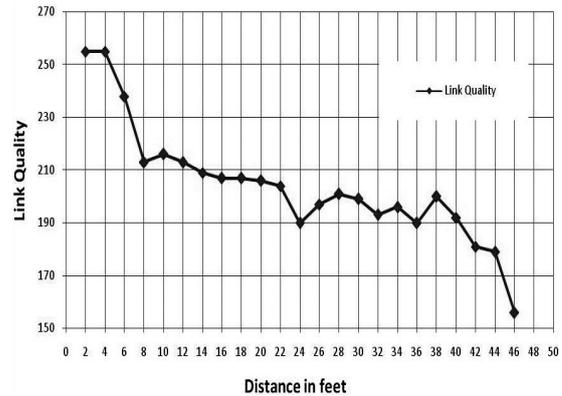

Fig.7 Distance vs. Link Quality

### E. Transmission bitrate vs. Power Consumption

It has been observed from Fig. 8 that power consumption of the bluetooth device decreases, when the bitrate of the audio stream is decreased.The power consumption for playing audio at various bitrates is calculated using Apple and Palm Pilot hardware.
.

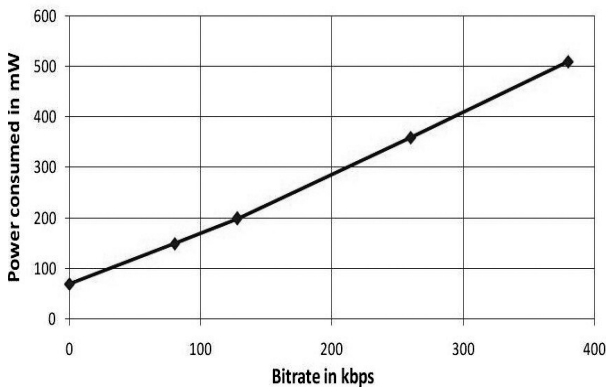

Fig.8 Bitrate vs. Power Consumption

## VI. CONCLUSION AND FUTURE WORK

Various Bluetooth signal parameters have been considered and their correspondence with distance has been analyzed. Based on analysis and experimental results, the following conclusions can be drawn:

1) Based on the analysis of the available Bluetooth signal parameters for the estimation of proximity, Link Quality has superior correlation with distance between the sender and receiver when compared to other signal parameters taken for study. Proximity estimation systems that rely on Link Quality would likely outperform any other proximity estimation system built upon other Bluetooth signal parameters.

2) Power consumption of the bluetooth device decreases, when the bitrate of the audio stream is decreased.

The directions for future work are as follows and we are working on it. It as been observed that the LQ readings do not vary much at close-range distances. On the other hand, the RSSI readings tend to change significantly at close-range distances. Therefore, a hybrid location system that combines both LQ and RSSI may be a viable option.

Normally, if a receiver measures an RSSI value that falls above the Golden Range (a positive value) then its Link Manager will request that the transmitter reduce its output power. If the RSSI value is below the Golden Range (a negative value), the Link Manager in the receiver will request that the transmitter increase its output power. The above mechanism is known as the feedback mechanism present in the Link Manager layer of the protocol. The values of the RSSI change reflect the output power. This makes the RSSI very fluctuating. If the feedback mechanism is disabled using hardware switches, then RSSI can be a more reliable parameter in the estimation of proximity.